\documentclass[aps,prl,twocolumn,superscriptaddress,footinbib]{revtex4-1}  
\usepackage{graphicx}  
\usepackage{dcolumn}   
\usepackage{bm}        
\usepackage{amssymb}   
\usepackage{amsmath}   
\usepackage{extarrows}
\usepackage{epstopdf}
\usepackage{hyperref}
\usepackage{color}

\newcommand{\hide}[1]{}
\newcommand{\Ho}{\hat{H}}
\newcommand{\rhoo}{\hat{\rho}}
\renewcommand{\ao}{\hat{c}}
\renewcommand{\aa}{\hat{c}^\dag}

\newcommand{\co}{\hat{c}}
\newcommand{\ca}{\hat{c}^\dag}
\newcommand{\no}{\hat{n}}
\newcommand{\ra}{\rangle}
\newcommand{\la}{\langle}

\newcommand{\br}{{\bm r}}

\newcommand{\up}{{\uparrow}}
\newcommand{\dn}{{\downarrow}}
\newcommand{\be}{\begin{equation}}
\newcommand{\ee}{\end{equation}}
\newcommand{\bes}{\begin{eqnarray}}
\newcommand{\ees}{\end{eqnarray}}

\definecolor{myblue}{rgb}{0,0,0.75}

\begin{document}
\author{Luis A. Pe\~na Ardila}
\email{luis@phys.au.dk}
\affiliation{Max Planck Institute for the Physics of Complex Systems,
N\"othnitzer Str.38, 01187,Dresden, Germany}
\affiliation{Institut for Fysik og Astronomi, Aarhus Universitet, 
8000 Aarhus C, Denmark}
\author{Markus Heyl}
\email{heyl@pks.mpg.de}
\affiliation{Max Planck Institute for the Physics of Complex Systems,
N\"othnitzer Str.38, 01187,Dresden, Germany}
\author{Andr\'e Eckardt}
\email{eckardt@pks.mpg.de}
\affiliation{Max Planck Institute for the Physics of Complex Systems,
N\"othnitzer Str.38, 01187,Dresden, Germany}

\date{\today}

\title{Measuring the single-particle density matrix for fermions and hard-core bosons in an optical lattice}

\begin{abstract}
Ultracold atoms in optical lattices provide clean, tunable, and well-isolated 
realizations of paradigmatic quantum lattice models. With the recent advent of 
quantum-gas microscopes, they now also offer the possibility to measure the
occupations of individual lattice sites. What, however, has not yet been achieved 
is to measure those elements of the single-particle density matrix, which are off-
diagonal in the occupation basis. Here, we propose a
scheme to access these basic quantities both for fermions as well as hard-core 
bosons and investigate its accuracy and feasibility. The scheme relies on the 
engineering of a large effective tunnel coupling between distant lattice sites 
and a protocol that is based on measuring site occupations after two 
subsequent quenches. 
\end{abstract}

\maketitle

\emph{Introduction.}---%
Atomic quantum gases in optical lattices \cite{BlochDalibardZwerger08,
LewensteinSanperaAhufinger, GrossBloch17} combine a variety of 
properties that make them a unique experimental platform for studying
mesoscopic quantum phenomena. 
%
Primarily, these systems provide clean realizations of paradigmatic quantum 
lattice Hamiltonians, well isolated from the environment.
%
Additionally, system parameters are highly adjustable including dimensionality, 
lattice geometry, interaction strengths, number of spin states, etc..
Moroever, optical lattice systems offer also unique measurement 
capabilities beyond what is possible in solid-state systems.
For example, using quantum-gas microscopes it is possible to measure the full 
spatial density profile with single-lattice-site resolution \cite{BakrEtAl09, 
ShersonEtAl10, CheukEtAl15, ParsonsEtAl15, HallerEtAl15, EdgeEtAl15, OmranEtAl15, 
Gross15, Kuhr16}. Upon repeating such experiments, one can determine means, 
fluctuations, correlations, and even full distribution functions of 
site occupations. This was used, e.g., to measure multi-particle string order 
\cite{EndresEtAl11,HilkerEtAl17}. 
Furthermore, combining these probes with measurement protocols, where additional dynamics is 
imposed, it has been achieved to detect entangled states of matter 
both by measuring a lower bound of the concurrence in spin systems~\cite{
MazzaEtAl15, Fukuhara2015} and by extracting Renyi entropies in
one-dimensional (1D) bosonic lattice systems~\cite{AlvesJaksch04,DaleyEtAl12, 
IslamEtAl15}.

Yet, these quantum-gas microscopes provide direct access only to physical 
quantities (near) diagonal in the occupation basis. 
While protocols for measuring currents and coherences on neighboring lattice sites
were proposed \cite{KilliEtAl12,KesslerMarquardt14}
and employed experimentally by either pairwise merging \cite{TrotzkyEtAl08, NacimbeneEtAl12,
GreifEtAl13} or isolating \cite{TrotzkyEtAl12} neighboring sites,
the off-diagonal matrix elements of the single-particle density matrix (SPDM)
\begin{equation}
	\chi_{\ell's',\ell s}= \langle \aa_{\ell's'}\ao_{\ell s}\rangle \, ,
\end{equation}
on distant non-neighboring lattice sites $\ell$ and $\ell'$ have not yet been accessed 
experimentally.
Here $\ao_{\ell s}$ denotes the annihilation operator for a particle with spin
$s$ on lattice site $\ell$.
The SPDM contains elementary information about the physical properties of quantum 
many-body systems and therefore is of interest on general grounds.
More specifically, one could use the SPDM to extract essential properties of
many-body localized phases \cite{BeraEtAl15, BeraEtAl17, LezamaEtAl17}, to detect 
topological Mott insulating states \cite{RaghuEtAl08}, or to probe topological 
edge states~\cite{HeEtAl17}. 
Moreover, it can be used to reconstruct the full reduced density matrix of two 
lattice sites. 
This would allow to access entanglement via the concurrence and the logarithmic
negativity~\cite{Plenio2014}, to study signatures of the butterfly effect~\cite{Chen17} 
or to detect many-body localized spin-glass order~\cite{Javanmard2018}.

\begin{figure}
\begin{center}
\includegraphics[width=1\linewidth]{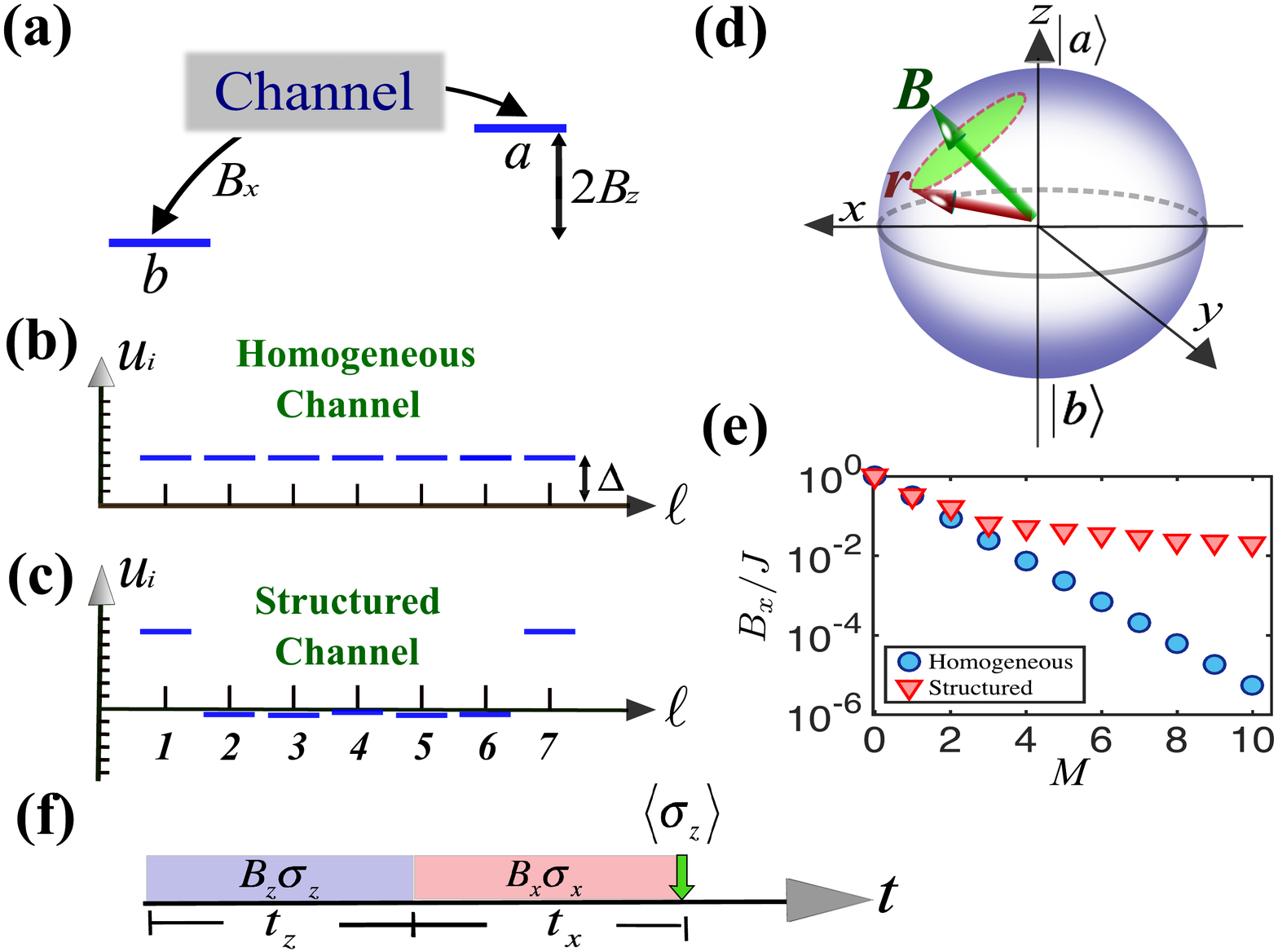}
\caption{(a) Two distant lattice sites $a$ and $b$ are effectively coupled to 
each other via energetically distant intermediate states of a channel formed by 
the sites between them. (b,c) Homogeneous and structured (optimized) channel design:
on-site potential $u_\ell$ on the sites $\ell=1,\ldots,M$ between $a$ and $b$, 
for $M=7$ and $\epsilon=0.05$. Horziontal axis defines zero energy 
$(u_a+u_b)/2\equiv 0$, tick marks are separated by $J$. (d) Rotation of the pseudospin
polarization $\br$. 
(e) Effective tunnel coupling $B_x$ vs.\ $M$ for $\epsilon=0.05$. (f) Measurement protocol.} 
\label{Fig1}
\end{center}
\end{figure}

Here we propose a feasible scheme for measuring the elements of the SPDM
$\chi_{\ell's',\ell s}$ for fermions or hard-core bosons. 
Our approach requires the use of two techniques as they are available
in the aforementioned quantum-gas microscopes:
single-site resolved density measurements and the ability to 
design high-resolution light-shift potentials using digital mirror devices.
Using these techniques we show (i) how two distant lattice sites $\ell'=a$ and
$\ell=b$ can be isolated from the rest of the system and efficiently coupled to
each other via virtual intermediate states forming a
``channel'' [Fig.~\ref{Fig1}(a)], and (ii) how to use this channel to measure
$\chi_{\ell's',\ell s}$ by monitoring only local densities on $a$ and $b$.
The main result concerning point (i) is depicted in Fig.~\ref{Fig1}(e). It shows 
that the effective parameter for tunneling between $a$ and $b$, $B_x$, whose 
inverse sets the measurement time scale, can be increased by orders of 
magnitude when structuring (optimizing) the potential landscape on the $M$
sites between $a$ and $b$ [Fig.~\ref{Fig1}(c)]. This channel optimization 
is absolutely crucial, since for unstructured (homogeneous) channels
[Fig.~\ref{Fig1}(b)] an exponential decay of $B_x$ with $M$ renders the 
measurement essentially impossible already for small distances. In turn,
for the structured channel $B_x$ scales like $1/M$ making measurements at 
longer distances much more feasible. In particular, this can allow for measuring
all relevant single-particle matrix elements of (many-body) localized systems, 
involving distances roughly up to the localization length.  
For problem (ii) we design a measurement protocol that is based on monitoring the
evolution of the densities on both sites after two subsequent quenches in the Hamiltonian.
It goes beyond other quench-based measurement schemes \cite{KilliEtAl12, OhlingerEtAl13, KesslerMarquardt14, 
HaukeEtAl14, WangEtAl17,GluzaEtAl18} that were implemented already successfully in experiment 
\cite{TrotzkyEtAl08, NacimbeneEtAl12, TrotzkyEtAl12, GreifEtAl13,AtalaEtAl14, FlaeschnerEtAl16, 
TarnowskiEtAl17}. 
In the following, we will first describe the measurement protocol (ii), then 
discuss the effective coupling of distant lattice sites (i), before closing with 
concluding remarks. 

\emph{\it Measurement protocol}---%
We will first treat the case of spinless fermions, before addressing the spinful 
problem and the case of hardcore bosons. The SPDM on two sites $a$ and
$b$ is then determined by:
\be
\chi_{a b}
    = \la\aa_{a}\ao_{b}\ra 
    = \mathrm{tr}\big\{\rhoo\aa_{a}\ao_{b}\big\} \, .
\label{eq:spdm}    
\ee
Here, $\rhoo$ denotes the system's full density operator, which can either 
describe an equilibrium state or can result from a nonequilibrium process.
As we show in more detail below, for our measurement protocol we engineer 
situations where the two sites $a$ and $b$ are isolated from the remainder 
of the system but still coupled to each other, such that their dynamics is
governed by two possible effective Hamiltonians:
\be
	\Ho_{ab}^z = B_z (\no_{a}-\no_{b}), 
    \quad \Ho^x_{ab}=B_x(\aa_{a}\ao_{b}+\aa_{b}\ao_{a}) \, ,
\label{eq:hoabc}
\ee
with $\no_\ell=\ca_\ell\co_\ell$, energy offset~$2B_z$, and
tunnel coupling~$B_x$. 

Let us first consider the subspace of a single fermion shared among both sites. 
It is convenient to introduce an effective pseudospin-$1/2$ representation with
$\up$ and $\dn$ referring to the cases where the fermion is located on site $a$ 
and $b$, respectively. 
Then we can recast the reduced density matrix into the form 
$\hat{\rho}_{ab} = [1+\br\cdot\hat{\bm \sigma}]/2$, with $\hat{\bm\sigma}
=(\hat{\sigma}_x,\hat{\sigma}_y,\hat{\sigma}_z)^\text{t}$ denoting the vector of 
Pauli matrices. It is characterized by the three-dimensional
Bloch vector $\br$ of length $|\br|\le 1$ directly corresponding to the 
polarization, $\la\hat{\bm \sigma}\ra=\mathrm{tr}\big(\hat{\rho}_{ab}
\hat{\bm \sigma}\big)=\br$. The purity reads
$\mathrm{tr}\big(\hat{\rho}^2_{ab}\big) =(1+\br^2)/2$, so that $|\br|<1$  
for mixed states. The SPDM $\chi_{ab}^{(1)}$ in the one-fermion subspace is 
given by 
\be\label{eq:chi1}
\chi_{ab}^{(1)}=\mathrm{tr}\big(\hat{\rho}_{ab}|\up\ra\la\dn|\big)
= \la\dn|\hat{\rho}_{ab}|\up\ra  = (r_x +ir_y)/2.
\ee 

%
Within the pseudospin representation, we can identify $B_z$ and $B_x$ as 
effective magnetic fields, i.e., $\Ho_{ab}^z = B_z \hat{\sigma}_z ,\quad \Ho_{ab}^x = B_x \hat{\sigma}_x $.
Let us now consider a protocol, where we first evolve the system with 
$\Ho_{ab}^z$ for a time $t_z$ and afterwards for a time $t_x$ with 
$\Ho_{ab}^x$. This amounts to two successive spin rotations [Fig.~\ref{Fig1}(d)]: 
one by the angle $\alpha=B_z t_z/(2\hbar)$ around the $z$ axis followed by one 
by the angle $\beta=B_x t_x/(2\hbar)$ around the $x$ axis. It transforms the 
polarization $\br$ before the rotation, which we wish to reconstruct, to the rotated
polarization $\br'=\br'(\alpha,\beta)$. Measuring the occupations $n_a$ and $n_b$ in
repeated experiments, one can obtain the $z$ polarization
$r_z'(\alpha,\beta)=\la\no_\up-\no_\dn\ra^{(1)}$ by averaging over the events with
$n=n_a+n_b=1$: 
\be
r_z'(\alpha,\beta) = \sin(\beta)[\sin(\alpha)r_x + \cos(\alpha)r_y]
                    +\cos(\beta)r_z.
\ee
The measurement protocol is depicted in Fig.~\ref{Fig1}(f). From measuring 
$r_z'(\alpha,\beta)$ for different angles $\alpha$ and $\beta$, we can 
reconstruct $\br$ [and, using Eq.~(\ref{eq:chi1}), also $\chi_{ab}^{(1)}$]:
\be
r_x = r_z'(\pi/2,\pi/2), \; r_y = r_z'(0,\pi/2), \; r_z=r_z'(0,0).
\ee

Note that the parameters $B_z$ and $B_x$, whose values control the angles
$\alpha$ and $\beta$, do not need to be known before the experiment, but can be 
measured from the periodicity of $r_z$ with respect to $t_z$ and $t_x$. 
Note also that, in case we can assume that $\hat{\rho}_{ab}$ 
describes a pure state, $|\br|=1$, we can reconstruct $r_y$, $r_z$, 
and $|r_x|=[1-r_y^2-r_z^2]^{1/2}$ without the need of implementing a finite $B_z$ 
for the $\alpha$ rotation. This has been exploited for the tomography of band 
insulators in momentum space \cite{HaukeEtAl14, FlaeschnerEtAl16}.

Let us now discuss the general case where a priori any particle number $n=n_a+n_b$ 
can occur.
When measuring the occupation numbers $n_{a}$ and $n_{b}$, one first has to 
distinguish between the three possible outcomes $n=0,1,2$, and note their 
relative frequencies $p_n$ in repeated experiments.
The cases $n=0$ and $n=2$ correspond to the states $|n_an_b\ra =|00\ra$ and
$|11\ra$, respectively, which are invariant under the action of both
Hamiltonians~(\ref{eq:hoabc}) and give 
$\chi_{ab}^{(0)}=\chi_{ab}^{(2)}=0$ of $\ca_a\co_b$. 
Overall, the full SPDM element can thus be written as 
\be
\chi_{ab} = \sum_{n=0}^2 p_n\chi_{ab}^{(n)} = p_1\chi_{ab}^{(1)}.
\label{eq:spdm2}   
\ee
The results obtained so far are equally valid for hard-core bosons, since
$\chi_{ab}^{(0)}=\chi_{ab}^{(2)}=0$ remains true and $\chi^{(1)}_{ab}$ is 
obtained from single-particle dynamics. The scheme does not generalize
to soft-core bosons.

Finally, it is left to consider spinful fermions. The SPDM for equal 
spin states $\la\aa_{a s}\ao_{b s}\ra$ can be obtained by repeating the above 
protocol individually for each spin state. This requires spin-sensitive 
measurements of $\no_{a s}$, as they were performed in various experiments. To 
obtain $\la\aa_{a s'}\ao_{b s}\ra$ for $s'\ne s$ one could first perform a spin 
rotation $s'\to s$ on site $\ell'$ and then, again, proceed as before. Since such 
a spin-rotation commutes with both quench Hamiltonians, it could be performed also 
after the double quench. However, this is in turn equivalent to simply measuring 
$\no_{s'a}$ instead of $\no_{s a}$, so that actually no spin rotation is needed. 
Note that unlike for spinless fermions, which are noninteracting, the implementation
of the quadratic Hamiltonians ~(\ref{eq:hoabc}) requires for spinful fermions also
to switch off the interactions using a Feshbach resonance \cite{ChinEtAl10}. This technique
is available in a fermionic quantum-gas microscope \cite{HilkerEtAl17}.

\begin{figure}
\begin{center}
\includegraphics[width=1\linewidth]{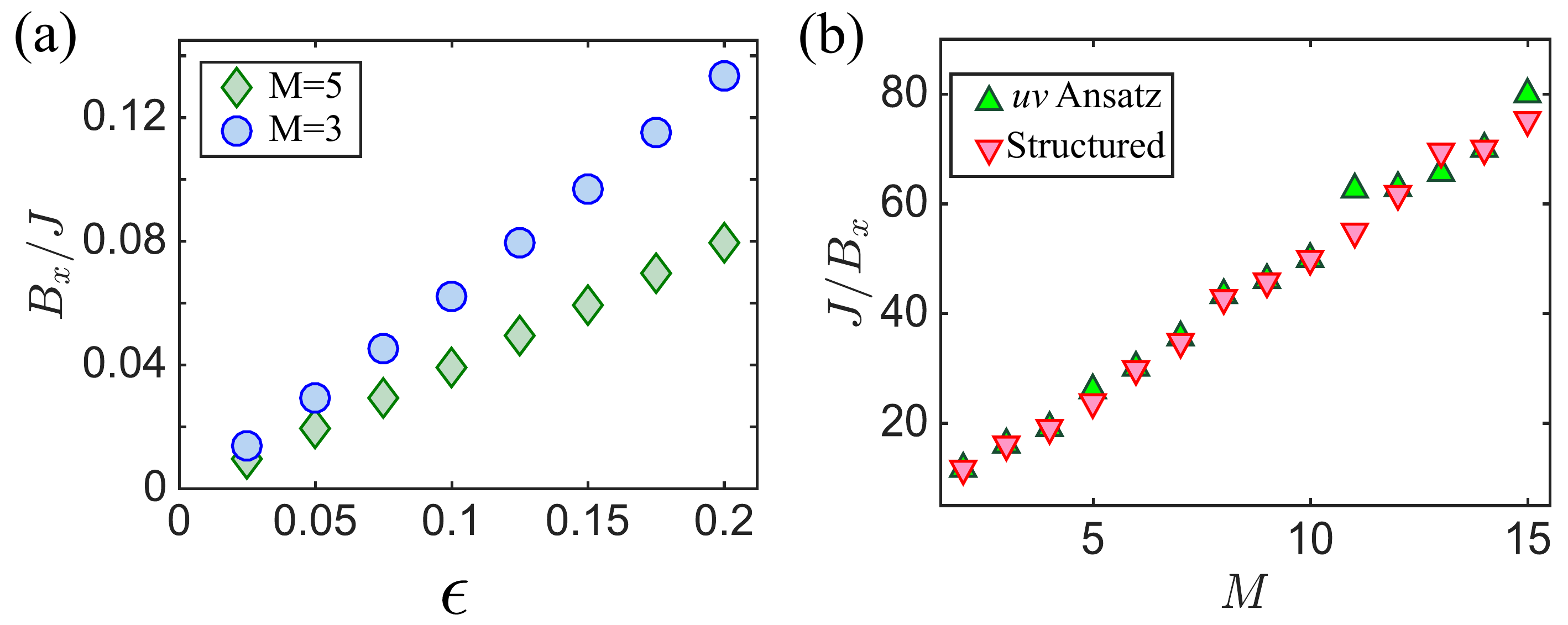} 
\caption{(a) Effective coupling versus the estimated error. (b) Inverse effective coupling versus channel length.}
\label{Fig2}
\end{center}
\end{figure}

\emph{Realizing the quench Hamiltonians.}---%
Let us now discuss how to implement the Hamiltonians $\Ho_{ab}^z$ and
$\Ho_{ab}^x$ experimentally. For simplicity, we will consider spinless fermions 
in 1D. The below reasoning applies also to hard-core bosons, which can 
be mapped to free fermions using a Jordan-Wigner transformation.
For the spinful case the same Hamiltonian has simply to be realized for all spin 
components. The generalization to higher spatial dimensions can be achieved 
by realizing a 1D channel connecting two sites or, alternatively,
by generalizing the below reasoning to higher-dimensional channel architectures.

We assume that during both stages of the measurement protocol the subsystem 
containing the two sites $a$ and $b$ as well as the $M$ sites between them 
(forming the channel) is decoupled from the rest of the lattice by 
switching on a large potential offset.  We label the lattice sites of this chain 
from $a$ ($\ell=0$) to $b$ ($\ell=M+1$) in ascending order:
\be
\Ho=-J\sum_{\ell = a}^{b-1} \left(\aa_{\ell}\ao_{\ell+1}+ \aa_{\ell+1}\ao_{\ell} \right) +\sum_{\ell=a}^b u_\ell \no_\ell  \, .
\ee
Here, $J$ denotes the tunneling amplitude and the $u_\ell$ are tunable
on-site energies. The latter can be tuned independently using 
high-resolution light-shift potentials realized by digital mirror devices
in quantum-gas microscopes.
For later convenience let us decompose this Hamiltonian according to
$\Ho= \Ho_{ab}+\Ho_\text{ch}+\Ho_\text{cp}$. Here $\Ho_{ab}=u_a\no_a+u_b\no_b$ 
captures the subspace containing the sites $\ell=a,b$. The channel between $a$ 
and $b$ is described by the Hamiltonian $\Ho_\text{ch}= -J\sum_{\ell = a+1}^{b-2} 
\big(\aa_{\ell+1}\ao_{\ell} + \text{h.c.}\big) 
+\sum_{\ell=a+1}^{b-1} u_\ell \no_\ell = \sum_k \alpha_k\ca_k\co_k $, which is
diagonalized by the modes $k=1,\ldots,M$, having energies $\alpha_k$ and 
annihilation operators $\co_k=\sum_{\ell=a+1}^{b-1}\lambda_{k \ell}\ao_\ell$
with real coefficients $\lambda_{k\ell}$. 
The channel is coupled to $a$ and $b$ by
$\Ho_\text{cp} = -J \big(\aa_a\ao_{1}+\aa_{b}\ao_{M} + \text{h.c.}\big) =
\sum_k \big(J_{ak}\aa_a\co_{k} + J_{bk}\aa_{b}\co_{k}+ \text{h.c.}\big)$,
where we defined the real matrix elements $J_{ak}=-J\lambda_{k(a+1)}$ and 
$J_{bk}=-J\lambda_{k(b-1)}$.

We aim at an effective Hamiltonian involving only the lattice sites $a$ and $b$ 
which we achieve by coupling them through the intermediate channel via virtual 
off-resonant processes. In this spirit, we consider a setup where 
the on-site energies $u_\ell$ are tuned in such a way that the tunneling from 
sites $a$ and $b$ into the channel is accompanied with a large energy cost. In 
this case, $\Ho_\text{cp}$ constitutes a weak perturbation, that can be 
eliminated using a Schrieffer-Wolff transformation: 
$e^{\kappa\hat{S}}[\Ho_{ab}+\Ho_\text{ch}+\kappa\Ho_\text{cp}]e^{-\kappa\hat{S}}
=\Ho_{ab}^\text{eff}+\Ho_\text{ch}^\text{eff}+\mathcal{O}(\kappa^3)$, with
counting parameter $\kappa=1$ and an antihermitian generator 
$\hat{S}= \sum_{k,x} \big(A_{xk}\aa_x\co_k -\text{h.c.}\big)$, which includes the small
parameters of our approximation, $A_{xk}\equiv J_{xk}/(\alpha_k-u_x)\ll 1$ for $x=a,b$. 
This gives the effective Hamiltonian
\be
\Ho_{ab}^\text{eff} = B_x (\aa_a\ao_b+\aa_b\ao_a) + B_z(\no_a-\no_b),
\ee
with $2B_x = -\sum_k (A_{ak}J_{bk}+A_{bk}J_{ak})$ and
$2B_z = u_a-u_b +\sum_k(A_{ak}J_{ak}-A_{bk}J_{bk})$, where we have dropped an 
irrelevant term $\propto (\no_a+\no_b)$. 
This Hamiltonian can take the form of both the desired quench Hamiltonians. For
$u_a=u_b=0$ one has $B_z=0$ and $B_x = -\sum_k J_{ak}J_{bk}/\alpha_k$ so that
$\Ho_{ab}^\text{eff}=\Ho_{ab}^x$. In turn, $\Ho_{ab}^\text{eff}\simeq\Ho_{ab}^z$ 
with $B_z=(u_a-u_b)/2$ can be achieved in the limit $A_{xk}\to 0$, which is reached
by increasing the channel energies $\alpha_k$ via potentials $u_\ell$. 


Within our scheme there are two sources of errors. First, we neglect corrections 
in the perturbative derivation of the effective Hamiltonian beyond second order. 
Second, the effective Hamiltonian is not realized in the basis of bare site 
occupations, but rather in the slightly rotated perturbed basis. The second error
is of second order (as we argue now) and therefore dominates. 
Considering $u_a=u_b=0$ as well as a reflection symmetric channel Hamiltonian 
(so that $A_{ak}^2=A_{bk}^2$) and assuming that terms containing off-diagonal expectation values
$\la\aa_a\co_k\ra$ and $\la\ca_{k'}\co_k\ra$ with $k'\ne k$ sum up to zero (rotating wave approximation), we find $\la e^{\hat{S}}(\no_a-\no_b)e^{-\hat{S}}\ra 
= (1-\epsilon)\la(\no_a-\no_b)\ra$, with the state-independent relative error 
between dressed and bare $z$ polarization 
 \be
 	\epsilon = \sum_k A_{ak}^2 \,.
 \ee
In the following we optimize our channel in such a way that we maximize $B_x$ 
while keeping the target error $\epsilon$ fixed. We also confirm numerically that the 
estimated error $\epsilon$ indeed quantifies deviations between the exact time evolution
and that generated by the effective Hamiltonian.

\begin{figure}
\begin{center}
\includegraphics[width=1\linewidth]{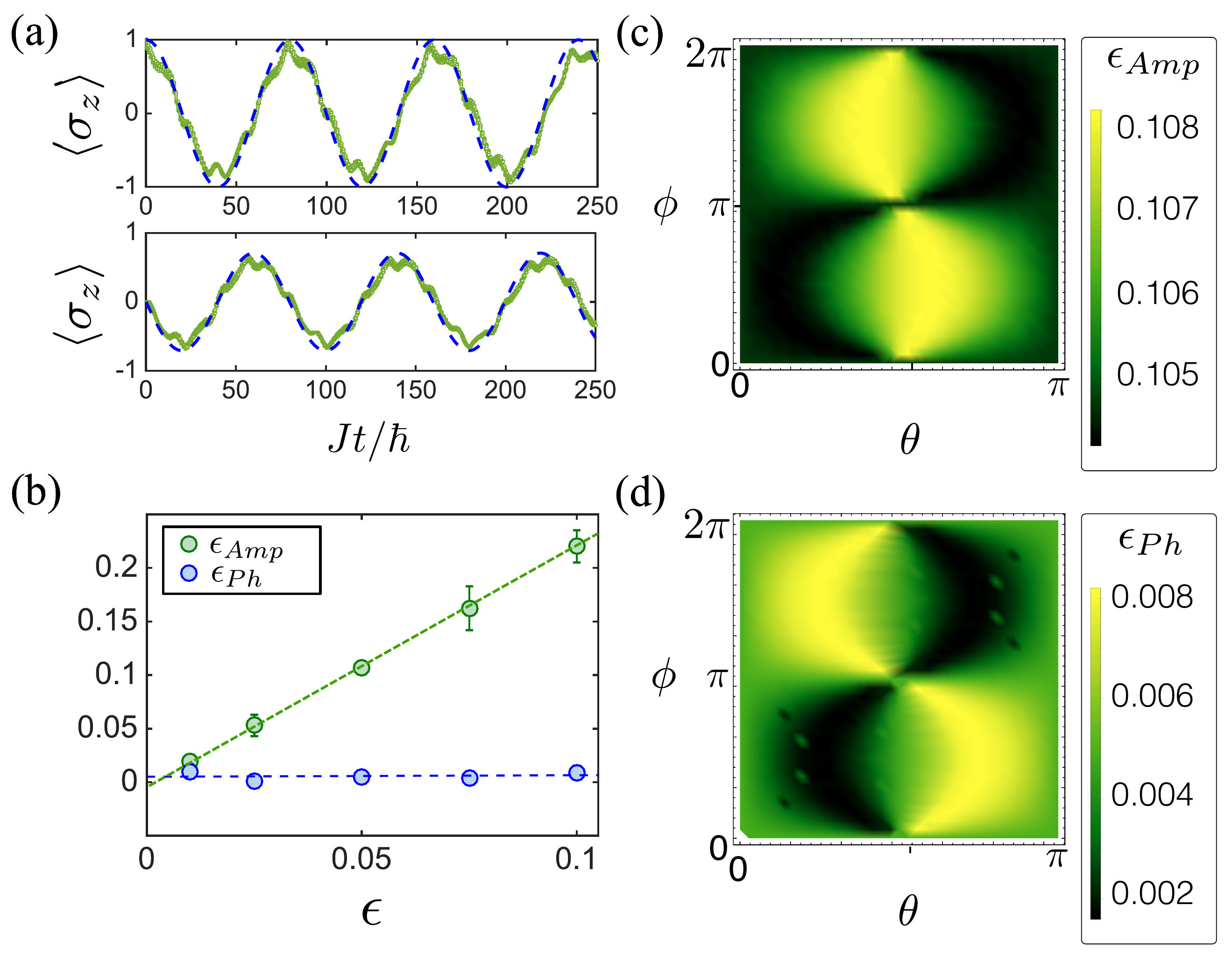} 
\caption{(a) Evolution of $z$ polarization for full and effective Hamiltonian 
(solid and dashed lines, respectively) for $\epsilon=0.05$,
and initial polarizations $(\theta,\phi)=(0,\pi/4)$ (upper) and $(\pi/2,\pi/4)$(lower). 
(b) Amplitude and phase error (averaged over initial polarization with fluctuations indicated by error bars) versus estimated error. 
(c,d) Amplitude and phase error versus initial polarization for $\epsilon=0.05$.
In all panels (a)-(d) $M=5$ and $|\br|=1$.}
\label{Fig3}
\end{center}
\end{figure}

\emph{Engineering the channel.}---%
Let us first consider a homogeneous channel [Fig.~\ref{Fig1}(b)] with
$u_\ell=\Delta\gg J$, giving rise to channel states with energies 
$\alpha_k=\Delta-2J\cos[\pi k/(M+1)]$ and wave functions
$\lambda_{k\ell}=[2/(M+1)]^{1/2}\sin[\pi k \ell/(M+1)]$ on the sites 
$\ell=1,\ldots, M$ of the channel (for convenience we assume $a=0$).
Choosing $\Delta$ for each $M$ so that $\epsilon=0.05$, we find an exponential
suppression of $B_x$ with $M$ [Fig.~\ref{Fig1}(e), circles], related to a 
destructive interference between channel modes. Namely $B_x$ is a sum over 
terms with oscillating sign, since $\lambda_{kM}=(-)^{k+1}\lambda_{k1}$. This 
unfavorable scaling restricts the scheme to short distances. 

To obtain better results, we investigate structured reflection-symmetric channel 
potentials $u_i$ [Fig.~\ref{Fig1}(c)]. We maximize $B_x$ again under the constraint
$\epsilon=0.05$ \footnote{For each set of channel parameters $u_i$, we diagonalize the 
channel Hamiltonian and compute both $B_x$ and $\epsilon$ using the expressions obtain
from perturbation theory.}.
{Fig1}(c) shows the optimized channel design for $M=7$. We
find a power-law instead, $B_x\sim 1/M$ [Fig.~\ref{Fig2}(b), triangles]. 
This tremendous improvement is one of the main result of this paper. It 
implies that significant effective coupling matrix elements can be engineered 
over rather large distances. Fig.~\ref{Fig2}(a) shows that the optimized coupling
$B_x$ scales proportional to the error.

Let us study the validity of our perturbative results and compare the evolution
of the $z$ polarization $\la(\no_a-\no_b)\ra$ generated by the full Hamiltonian
$\Ho_{ab}+\Ho_\text{ch}+\Ho_\text{cp}$ to the one obtained from the effective 
Hamiltonian $\Ho_{ab}^\text{eff}=\Ho_{ab}^x$. Two examples are shown in
Fig.~\ref{Fig3}(a). Fitting $A\sin(\omega t-\eta)$ to the exact 
evolution, we can compare the extracted amplitude $A$ and phase $\eta$ ($\omega$ 
to the perturbative results $A_\text{eff}$ and $\eta_\text{eff}$. In
Fig.~\ref{Fig3}(c) and (d) we plot the errors $\epsilon_\text{Amp}=|A-A_\text{eff}|/|A|$ and 
$\epsilon_\text{Ph}=|\eta-\eta_\text{eff}|$ versus the initial polarization
$\br$, characterized by $|\br|=1$ and polar angles $(\theta,\phi)$. Their 
behavior with respect to the estimated error $\epsilon$ is depicted in
Figs.~\ref{Fig3}(b). The amplitude error $\epsilon_\text{Amp}$ is directly 
related to the estimated error $\epsilon$. Like $\epsilon$ it (practically)
does not depend on the state [see limits of color bar in Fig.~\ref{Fig3}(c)] and
we find a linear scaling $\epsilon_\text{amp}\approx 2 \epsilon$ [Fig.~\ref{Fig3}(b)], 
consistent with the fact that projecting between bare and the perturbeed basis states
leads to an error twice (when switching on $\Ho_{ab}^{x}$ and when recording
bare site occupations afterwards). 
The phase error is much smaller [$\lesssim 10^{-2}$, see Fig.~\ref{Fig3}(d)] and
does not show a strong dependence on $\epsilon$. This justifies a posteriori our
choice to quantify the error via $\epsilon$.

Finally, we wish to shed light on the physics underlying the $1/M$ scaling of
$B_x$. For this purpose, we consider a simple model for a structured channel 
and show that it gives rise to such behavior. Motivated by the results
of the channel optimization (see Fig.~{Fig1}(c) for the $M=7$) , we define 
$u_1=u_M=u$ and $u_{2}=\cdots=u_{M-1}=v$. The energy $v$ shall be comparable 
to $J$, while $u$ takes a large value forming a barrier at both channel edges. In
this way tunneling from $a$ ($b$) into the central channel ($i=2,\ldots, M-1$) via the 
edge site $1$ ($M$) can be viewed as a second-order process with effective matrix 
element $-J_\text{eff}\approx J^2/u$. Diagonalizing the homogeneous central 
channel, we obtain modes $q=1,\ldots, M-2$ with energies
$\beta_q= v-2J\cos[\pi q/(M-1)]$. The maximum energy separation from neighboring 
levels $\Delta\beta\approx2\pi J/M$ is found for modes $q\approx(M-1)/2$; let
$q_0$ be one of them. By tuning $u$ and $v$, we can achieve that
$|J_\text{eff}\lambda| \ll |\epsilon_{q_0}| \ll \Delta\beta$, where
$\lambda \approx \sqrt{2/M}$ and $(-)^{q_0+1}\lambda$ are the overlaps of mode
$q_0$ with site $i=2$ and $i=M-1$, respectively. 
Let's say, $\epsilon_{q_0} =\delta\Delta\beta$ and
$J_\text{eff}\lambda =\delta^2\Delta\beta$, with small parameter $|\delta|\ll1$.
This corresponds to a situation, where both sites $a$ and $b$ couple to each 
other predominantly through a single central channel mode $q_0$, so that 
destructive interference between channel modes is avoided. Now, we can estimate
$B_x$ as resulting from a second-order process on a next level, connecting $a$ 
and $b$ via the intermediate virtual state $q_0$. We find 
$B_x\approx (-)^{q_0} \lambda^2 J_\text{eff}^2/\epsilon_{q_0} \propto J/M$. 
Computing $B_x$ within this ansatz by optimizing $u$ and $v$ for each 
system size $M$, keeping $\epsilon=0.05$ fixed, clearly confirms the linear scaling
of $J/B_x$ with $M$ [Fig.~\ref{Fig2}(b), green triangles]. Surprisingly, the $uv$ ansatz
works almost as good as the full channel optimization (red triangles), which also 
give large barrier potentials $u_1=u_M$. Thus, apart from providing insight into the 
mechanism underlying the channel architecture, the $uv$ ansatz provides also a simple
recipe for  the experimental implementation of our measurement scheme. 

\emph{Conclusions.}---%
In summary, we have proposed an experimental scheme for measuring off-diagonal elements
of the SPDM in lattice-site representation for fermions and hard-core bosons in optical
lattices. It relies on the ability of quantum-gas microscopes to both measure occupations
and create light-shift potentials with single-site resolution. For this purpose we showed
on the one hand, how to engineer a significant effective tunnel coupling between distant 
lattice sites. On the other hand, we presented a protocol that uses the dynamics induced 
by suddenly switching on this coupling for reconstructing the sought-after matrix elements
between two sites $\ell$ and $\ell'$ from measuring the occupations on these sites only.


\begin{acknowledgments}
Andr\'e Eckardt acknowledges support from the Deutsche Forschungsgemeinschaft (DFG) via the Research Unit FOR 2414 (grant number EC 392/3-1) and Markus Heyl by the Deutsche Forschungsgemeinschaft via the Gottfried  Wilhelm  Leibniz  Prize  program.
\end{acknowledgments}

\bibliographystyle{apsrev4-1}
\bibliography{mybib}

\end{document}